\documentclass{article}

     \usepackage[preprint]{neurips_2019}

\usepackage{graphicx}
\usepackage{subfigure}
\usepackage{chessboard}
\usepackage{afterpage}
\storechessboardstyle{10x10}{maxfield=j10,labelleftwidth=1.5ex,showmover=false,labelleft=false,labelbottom=false}
\storechessboardstyle{6x6}{maxfield=f6,showmover=false,labelleft=false,labelbottom=false,pgfstyle=straightmove,arrow={}}
\usepackage{amsmath,amsfonts,amssymb}
\usepackage{tabularx,supertabular,booktabs,lscape,array,longtable}
\newcolumntype{R}[1]{>{\raggedleft\let\newline\\\arraybackslash\hspace{0pt}}m{#1}}
\newcolumntype{L}[1]{>{\raggedright\let\newline\\\arraybackslash\hspace{0pt}}m{#1}}
\usepackage[vlined,linesnumbered,ruled]{algorithm2e}  
\usepackage{url}
\usepackage[bookmarks,implicit=true,colorlinks,linkcolor=black,citecolor=black,urlcolor=blue]{hyperref} 

\def\cplex{\texttt{ IBM-ILOG CPLEX Cplex 12.7.1}}   

\usepackage{color}

\newcommand{\none}[1]{{#1}}
\def\matteo{\none}
\def\domenico{\none}
\def\matt{\none}
\def\matte{\none}

\title{Finding First and Most-Beautiful Queens \\ by Integer Programming}

\author{
  Matteo Fischetti \thanks{corresponding author: \texttt{http://www.dei.unipd.it/\textasciitilde fisch}} \\
  Department of Information Engineering\\
  University of Padova, Italy\\
  \texttt{matteo.fischetti@unipd.it} \\
  \And
  Domenico Salvagnin \\
  Department of Information Engineering\\
  University of Padova, Italy\\
  \texttt{domenico.salvagnin@unipd.it} 
}

\begin{document}

\maketitle
    
\begin{abstract}
The $n$-queens puzzle is a well-known combinatorial problem that requires to place $n$ queens on an $n\times n$ chessboard so that no two queens can attack each other. Since the 19th century, this problem was studied by many mathematicians and computer scientists. While finding any solution to the $n$-queens puzzle is rather straightforward, it is very challenging to find the lexicographically first (or smallest) feasible solution. Solutions for this type are known in the literature for $n\le 55$, while for some larger chessboards only partial solutions are known. The present paper was motivated by the question of whether Integer Linear Programming (ILP) can be used to compute solutions for some open instances. We describe alternative ILP-based solution approaches, and show that they are indeed able to compute (sometimes in unexpectedly-short computing times) many new lexicographically optimal solutions for $n$ ranging from $56$ to $115$. 
\matt{One of the proposed algorithms is a pure cutting plane method based on a combinatorial variant of classical Gomory cuts.} 
\matteo{We also address an intriguing ``lexicographic bottleneck'' (or min-max) variant of the problem that requires finding a most beautiful (in a well defined sense) placement, and report its solution for $n$ up to 176.}
\end{abstract}

~

\section{Introduction}\label{sec:intro}

The $n$-queens puzzle is a well-known combinatorial problem that requires to place $n$ queens on an $n\times n$ chessboard so that no two queens can attack each other, i.e., no two queens are on the same row, column or diagonal of the chessboard. Initially stated for the regular $8 \times 8$ chessboard in 1848~\cite{Bezzel}, it was soon generalized to the $n \times n$ case~\cite{Lionnet}, and has attracted the interest of many mathematicians (including Carl Friedrich Gauss) and, more recently, by Edsger Dijkstra who used it to illustrate a depth-first backtracking algorithm. As a decision problem, the $n$-queens puzzle is rather trivial, as a solution exists for all $n > 3$, and there are closed formulas to compute such solutions; see, e.g.,  the survey in~\cite{BellStevens}. On the other hand, the counting version of the problem, i.e., to determine the number of different ways to put $n$ queens on a $n\times n$ chessboard turns out to be extremely challenging. The sequence, labelled \texttt{A000170} on the Online Encyclopedia of Integer Sequences (OEIS)~\cite{OEIS}, is currently known only up to $n=27$. The related problem of finding all solutions to the problem was shown in~\cite{HHS} to be beyond the \#P-class.

Another variant of the problem, which is somewhat related to the one addressed in this paper, is the $n$-queens completion problem, in which some queens are already placed on the chessboard and the solver is required to place the remaining ones, or show that it is not possible. The $n$-queens completion problem is both NP-complete and \#P-complete, as proved in~\cite{GJN}.

Following a suggestion of Donald Knuth \cite{don}, in this paper we study another very challenging version of the $n$-queens problem, namely, finding the lexicographically-first (or smallest) feasible solution. This is sequence \texttt{A141843} on OEIS. Solutions for this variant are known only for $n\le 55$ \cite{web}, while for some larger chessboards only partial solutions are known.

It is worth noting that the lexicographically optimal solution is known for the case of a chessboard of infinite size. Indeed, such a sequence can be easily computed by a simple greedy algorithm that iterates over the anti-diagonals of the chessboard and places a queen in each anti-diagonal in the first available position (this is sequence \texttt{A065188} on OEIS). Interestingly, as the size of the chessboard increases, its lexicographically optimal solution overlaps more and more with this greedy sequence.   

\matteo{
Finally, we address a very intriguing variant of the problem, also proposed to us by Donald Knuth \cite{don18}. This variant calls for a solution where the queens are placed so as to minimize the multiset of distances to the center of the board. This solution (which is not unique) enjoys  a number of nice properties (including double symmetry) and was argued to be the ``most-beautiful'' placement of the queens in a blackboard.

To be more specific, Knuth proposed the following {\em lexicographic bottleneck} (or min-max) variant of a classical lexicographic optimization problem: 
given a ground set of available options and the associated {\em costs}, find a feasible solution w.r.t. to a given set of constraints that minimizes (lexicographically) its maximum cost, and then the second-maximum, and so on. At first glance, this problem can be solved  first sorting the options in non-increasing order of the associated costs, and then by finding the corresponding lexicographic minimal feasible solution. If the costs are all different, this approach is indeed correct and produces the required min-max optimal solution. When repeated costs are allowed, however, different orderings of the costs can lead to very different (suboptimal) final solutions and the approach, as stated, is wrong---hence a more clever approach has to be applied. This latter situation arises, in particular, in the {\em most-beautiful $n$-queens problem} where the options are the blackboard positions, and the costs measure the distance of each cell from the center of the chessboard. Solutions for this problem are only known for $n$ up to 48 \cite{don18}.
}

The outline of the paper is as follows. In Section~\ref{sec:basicmodel} we describe the basic Integer Linear Programming (ILP) formulation for the $n$-queens model, as well as potential families of valid inequalities. In Section~\ref{sec:solution} we describe the different methods developed to solve the instances to lexicographic optimality, \matteo{and computationally compare them in Section~\ref{sec:comp}. In Section~\ref{sec:most-beautiful} we show how to solve a lexicographic bottleneck problem (and, in particular, the most-beautiful $n$-queens problem) through a sequence of ILPs.}                    
Conclusions and future directions of research are drawn in Section~\ref{sec:conclusions}. Finally, we list in Appendix all the new \matteo{lexicographically-first} solutions we found for $n$ ranging from $56$ to $115$, \matteo{and also report the most-beautiful solutions for some values of $n$ up to 176.}     

\matteo{A preliminary version of the present paper was presented at the international conference on the Integration of Constraint Programming, Artificial Intelligence, and Operations Research (CPAIOR) held in Delft, The Netherlands, on 5-8 June, 2018 \cite{noi}.}

\section{An ILP model}
\label{sec:basicmodel}

A basic ILP model for the $n$-queens problem can be obtained by introducing the binary variables $x_{ij}=1$ iff a queen is placed in row $i$ and column $j$ of the chessboard, for each $i,j=1,\dots,n$. Constraints in the basic model stipulate that (i) there is exactly one $x_{ij}=1$ in each row $i$; (ii) there is exactly one $x_{ij}=1$ in each column $j$; and (iii) there is at most one $x_{ij}=1$ in each diagonal of the chessboard. Note that all such constraints are \emph{clique} constraints.

In principle, it would be possible to encode the (row-wise) lexicographically minimum requirement by just adding the objective function:
\begin{align}
	\sum_{i=1}^n \sum_{j=1}^n 2^{ni+j} x_{ij} \label{eq:lexobj}
\end{align}
and solve the problem with a black-box ILP solver. However, the size of the coefficients makes such a method practical only for the smallest chessboards. Still, this simple model, without the objective \eqref{eq:lexobj}, is the basis of all the methods that will be discussed in Section~\ref{sec:solution}.

\begin{figure}
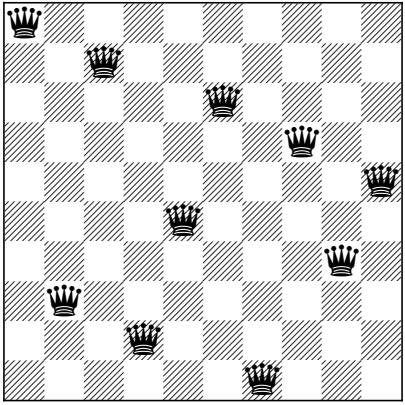

\begin{center}
\chessboard[style=10x10,,smallboard,setblack={Qg1,Qd2,Qb3,Qi4,Qe5,Qj6,Qh7,Qf8,Qc9,Qa10}]
\caption{Lexicographically optimal solution for $n=10$.}
\label{fig:q10}
\end{center}
\end{figure}

A compact way to represent a feasible solution is to use a permutation $\pi=(\pi_1, \dots, \pi_n)$ of the integers $1,\dots,n$ defined as follows:
\begin{align}
	\pi_i := \sum_{j=1}^n j \, x_{ij}, \ \ \ \ i=1,\dots,n. \label{eq:map}
\end{align}
Among all permutations $\pi$ that correspond to a feasible $x$, we then look for the lexicographically smallest one.
For example, the lex-optimal solution for $n=10$, depicted in Figure~\ref{fig:q10}, can be described as
\[(1, 3, 6, 8, 10, 5, 9, 2, 4, 7).\]

The $n$-queens problem can also be easily reformulated as a maximum independent set problem, as noted for example in~\cite{FouldsJohnston}. Indeed, one just needs to construct a graph in which there is a node for each square of the chessboard and an edge for each pair of conflicting squares, i.e., for any two squares in the same row, column or diagonal. Then any independent set of cardinality $n$ is a solution to the puzzle. The independent set reformulation immediately suggests classes of valid inequalities for the $n$-queens problem, namely all that are valid for the stable set polytope, such as \emph{clique} and \emph{odd-cycle}~\cite{GLS88} inequalities.

\begin{figure}
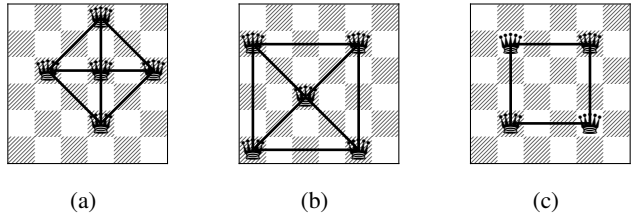

\begin{center}
\subfigure[]{
\chessboard[style=6x6,
	tinyboard,
	setblack={Qd2,Qb4,Qd4,Qf4,Qd6},
	markmoves={b4-d6,b4-d4,b4-d2,d4-d6,d2-d4,d4-f4,d2-f4,d6-f4}]
\label{fig:clique1}}
\subfigure[]{
\chessboard[style=6x6,
	tinyboard,
	setblack={Qa1,Qe1,Qc3,Qa5,Qe5},
	markmoves={a1-e1,a5-e5,a1-a5,e1-e5,c3-a1,c3-e1,c3-a5,c3-e5}]
\label{fig:clique1a}}
\subfigure[]{
\chessboard[style=6x6,
	tinyboard,
	setblack={Qb2,Qe2,Qb5,Qe5},
	markmoves={b2-e2,e2-e5,b5-e5,b2-b5}]
\label{fig:clique2}}
\caption{Three different families of clique cuts for $n$-queens.}
\label{fig:cliques}
\end{center}
\end{figure}

Among clique inequalities, the following (polynomial in $n$) family is particularly relevant for our problem:
\begin{align}
x_{ij} + x_{i,j+h} + x_{i+h,j} + x_{i-h,j} + x_{i,j-h} &\le 1 \label{eq:clique1}  \\
x_{ij} + x_{i+h,j+h} + x_{i-h,j+h} + x_{i-h,j-h} + x_{i+h,j-h} &\le 1 \label{eq:clique1a} \\
x_{ij} + x_{i+h,j} + x_{i+h,j+h} + x_{i,j+h}  &\le 1 \label{eq:clique2}
\end{align}
where $i,j,h\in \{1,\dots, n\}$; of course, variables $x_{uv}$ corresponding to a position $(u,v)$ outside the $n \times n$ chessboard are removed from the summations. The three different types of cliques in this family are depicted in Figure~\ref{fig:cliques}.

Clique inequalities \eqref{eq:clique1}--\eqref{eq:clique2} can be trivially separated in time that is polynomial in $n$. In addition, in preliminary experiments we implemented a general-purpose exact clique separator based on the solution of an auxiliary ILP model, and it never produced any additional violated clique inequality for the instances in our testbed.

\begin{figure}
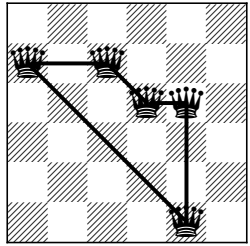

\begin{center}
\chessboard[style=6x6,
	smallboard, 
	setblack={Qa5,Qc5,Qd4,Qe4,Qe1},
	markmoves={a5-c5,c5-d4,d4-e4,e4-e1,e1-a5}]
\caption{Example of odd-cycle inequality for $n$-queens: no more than two of the five positions can be occupied by a queen.}
\label{fig:oddcycle}
\end{center}
\end{figure}

A second class of inequalities contains the so-called \emph{odd-cycle} inequalities. Given any odd cycle $O$ in the graph, the following inequality:
\begin{equation}
\sum_{k \in O} x_k \le \frac{|O|-1}{2}
\end{equation}
is valid for the stable set polytope. Odd-cycle inequalities can be easily separated as $\{0,1/2\}$-cuts with the combinatorial procedures described in~\cite{CapraraFischetti96,CapraraFischetti96b,Andreello2007}. An example of odd-cycle inequality occurring in the $n$-queens problem is illustrated in Figure~\ref{fig:oddcycle}.

\section{Solution methods}
\label{sec:solution}  

We next describe the solution algorithms that we implemented. 

\subsection{Using a Constraint Programming solver}
\label{sec:CP}

The $n$-queens puzzle can be easily modeled as a Constraint Programming (CP) problem. Indeed, working directly on the variables $\pi_i$, the puzzle can be formulated by just three \texttt{alldifferent}~\cite{Regin94,VanHoeve01} global constraints:
\begin{align}
\mathtt{alldifferent}(\pi_i, i=1,\dots,n) \label{eq:alldiff1} \\
\mathtt{alldifferent}(\pi_i + i, i=1,\dots,n) \label{eq:alldiff2} \\
\mathtt{alldifferent}(\pi_i - i, i=1,\dots,n). \label{eq:alldiff3}
\end{align}
We implemented the model above with Gecode~\cite{Gecode}. In order to enforce the model to find the lexicographically-smallest solution, we use Depth-First Search (DFS) as search strategy, always branching on the first unfixed variable $\pi_i$ and picking values in increasing order---in Gecode terminology, that amounts to using a \emph{brancher} specified by \texttt{INT\_VAR\_NONE()} and \texttt{INT\_VALUES\_MIN()}. In the following, we will refer to this solution method as \texttt{CP}.

\subsection{Using an exact ILP solver}
\label{sec:exact}

A simple algorithm to compute the lex-optimal solution by iteratively using a black-box ILP solver is as follows: We scan all the chessboard positions $(i,j)$ in lexicographical order, i.e., row by row. For each $(i,j)$, we are given the queens already positioned in the previous iterations (i.e., we have a number of fixed $x$ variables), and our order of business is to decide whether a queen can be placed in $(i,j)$ or not. This in turn requires solving the basic ILP model with some variables fixed in the previous iterations, by \emph{maximizing} $x_{ij}$: if the final optimal solution has value 1, we place a new queen in position $(i,j)$ by fixing $x_{ij}=1$, otherwise we fix $x_{ij}=0$ and proceed with the next chessboard position\footnote{Alternatively, one could fix $x_{ij}=1$, check the resulting model for feasibility, and then move to the next position.}. This approach requires solving $n^2$ ILPs.

In our actual implementation, a more effective scheme is used that exploits representation \eqref{eq:map}. To be specific, we scan the rows $i=1,\dots,n$, in sequence. For each $i$, we have already fixed in the previous iterations the lex-optimal sequence $\pi_1,\cdots,\pi_{i-1}$ and the corresponding $x$ variables, and we want to compute the \emph{smallest} feasible integer $\pi_i$. To this end we solve the basic ILP model, with some variables fixed in the previous iterations, by \emph{minimizing} the objective function \eqref{eq:map}, fix all the $x_{ij}$ variables in row $i$ accordingly, and proceed with the next row. In this way, only $n$ ILPs need to be solved. In the following, we will refer to this solution method as \texttt{ILP-ITER}.

\subsection{Using a truncated ILP solver}
\label{sec:heu}

We also implemented an explicit depth-first backtracking algorithm to build the lex-optimal permutation $\pi$, very much in the spirit of the CP approach described in Subsection~\ref{sec:CP}. At each iteration (i.e., at each node of the branching tree) we have tentatively fixed a lex-minimal, but possibly infeasible sequence,  $(\overline\pi_1, \dots,\overline\pi_{i-1})$  and the corresponding $x$ variables, and we have to decide the next value in position $i$. This is in turn obtained by solving a \emph{relaxation} of the current ILP with objective function \eqref{eq:map}, to be minimized, i.e., by applying the following three steps:
\begin{itemize}
	\item[i)] invoke the ILP solver (with its default cutting-plane generation and preprocessing) for a limited number of nodes, say $NN$;
	\item[ii)] define $\overline \pi_i$ as the best \emph{lower bound} available at the node limit (rounded up);
	\item[iii)] tentatively fix $\overline\pi_i$, along with the corresponding $x$ variables, as the $i$-th value in the sequence.
\end{itemize}
As a lower bound (instead of the true value) is used, it may happen that, at a later iteration, the current ILP becomes infeasible, proving that the current tentative subsequence $(\overline\pi_1,\cdots,\overline\pi_k)$ till position $k$ (say) is infeasible as well. In this case, a backtracking operation takes place, that consists in imposing that the $k$-th position must hold a value strictly larger than $\overline\pi_k$. The latter requirement can easily be enforced in the ILP model by setting $x_{kj}=0$ for $j=1,\dots, \overline\pi_k$.The algorithm ends as soon as the first feasible complete permutation $(\overline\pi_1, \dots,\overline\pi_n)$ is found.

After some preliminary tests, we decided to set $NN=0$, i.e., to only solve the root node of the ILP at hand. Note that this is \emph{not} equivalent to solving the LP relaxation of the ILP, as cutting planes and (most importantly) preprocessing play a crucial role here. According to our computational experience, solving just the LP relaxation is indeed mathematically correct and very fast, as the dual simplex can be used to reoptimize each LP, but the number of backtrackings becomes too large to have a competitive implementation. In the following, we will refer to this solution method as \texttt{ILP-TRUNC}.

\subsection{An enumerative method based on lexicographic simplex}
\label{sec:lexdfs}

Finally, given the strong lexicographic nature of the problem at hand, we decided to implement a custom enumerative algorithm based on the lexicographic simplex method~\cite{Gomory58,Gomory60}. The lexicographic simplex method not only finds an optimal solution to a given LP, but it guarantees to return the lexicographically smallest (or greatest) one among all optimal solutions. The lexicographic variant of the simplex method can be implemented quite easily on top of a black-box regular simplex solver, as described for example in \cite{BFZ10,ZFB11}. The idea is as follows. Given an ordered sequence of objective functions $f_k$ to optimize lexicographically, at each step we impose to stay on the optimal face of the current objective by fixing all variables (including the artificial variables associated to inequality constraints) with nonzero reduced cost, move to the next objective and reoptimize. Once all objectives have been optimized, in sequence, the original bounds for all variables are restored, which does not change the optimality status of the final basis, which is the lex-optimal one.

In our $n$-queens case, given our encoding of the permutation variables $\pi$ as $x_{ij}$, we are interested in the lexicographically maximal solution in the $x$ space or, equivalently, the sequence of objective functions to be minimized is $-x_{ij}$, for all $i,j=1,\dots,n$.

Using a lexicographic simplex method within an enumerative DFS scheme, in which again we always branch on the first unfixed variable and explore the $1$-branch first, provides the following advantages over using a ``regular'' simplex method:
\begin{itemize}
\item Whenever the LP relaxation turns out to be integer, i.e., there are no fractional variables, we are guaranteed that this is the lex-optimal integer solution within the current subtree, hence we can prune the node. Given our branching and exploration strategy, this also implies that we are done.
\item If the first unfixed variable at the current node gets a value strictly less than one, then we can fix the variable to zero.
This is easily proved using the lex-optimality of the LP solution as an argument. Being the first unfixed variable, this is the first objective to be considered by the lexicographic simplex at the current node, so a lex-optimal value $<1$ means that there is no feasible solution (in the current subtree) in which this variable takes value $1$. Note that this reduction can be applied iteratively until the first unfixed variable gets a value of $1$. We call this process \emph{mini-cutloop}.
\end{itemize}

The basic scheme above can be improved with some additional modifications. First of all, we do not need to branch on single variables but we can branch directly on rows, again always picking the first row that contains an unfixed variable. For example, let the first unfixed variable be $x_{ij}$: instead of branching on the binary dichotomy
$x_{ij} = 1 \vee x_{ij} = 0$, we use the $n$-way branching $x_{i1} = 1 \vee x_{i2} = 1 \vee \ldots \vee x_{in} = 1$. Of course, variables that are already fixed are removed from the list. This basically mimics the branching that would have been done by working directly with the $\pi$ variables, as done by the CP solver.

Note that, because of our rigid branching strategy, there is no need for a full lexicographic optimization at each node. Indeed, for the purpose of branching, we can stop the lexicographic optimization at the first fractional variable, as we will be forced to branch on its row, or on a previous one. For this very reason, and because of the $n$-queens structure, we implemented a specialized lexicographic simplex method, where instead of optimizing one variable at the time, we optimize row by row, also integrating the mini-cutloop in the process. In particular, we do the following:
\begin{enumerate}
\item Let $i^*$ be the first row with an unfixed variable. Set the objective function to $\sum_{j=1}^n j x_{i^*j}$ and minimize it.
\item Apply the mini-cutloop, by iteratively fixing the first unfixed variable in the row if its fractional value is $<1$ and by reoptimizing with the dual simplex.
\item If all variables in the current row are fixed this way, then we can move to the next row and go to step (1). Otherwise stop.
\end{enumerate}
Note that the method above does not need to temporarily fix variables as the regular lexicographic simplex would. It is also important to note that, in the loop above, if the current fractional solution is integer, we are no longer guaranteed that this is the lexicographically optimal solution. In this (rare) case, we resort to a full-blown lexicographic simplex method to tell whether we can prune the node or need to branch.

The effectiveness of the node processing above greatly depends on the mini-cutloop, which in turn relies on being able to recognize fixed variables, i.e., to distinguish between a variable that happens to be zero or one in the current fractional solution, and a variable that is actually fixed at that value in the current node. For this purpose, we implemented a specialized propagator for the clique constraints of the basic model---while there is no need to propagate the clique constraints (\ref{eq:clique1})--(\ref{eq:clique2}) as those can never lead to additional fixings.

Finally, separation of the clique inequalities (\ref{eq:clique1})--(\ref{eq:clique2}) and odd-cycle inequalities has also been implemented and added to the node processing code. In the following, we will refer to this solution method as \texttt{LEX-DFS}.

\domenico{
\subsection{A pure cutting plane method based on lexicographic simplex}
\label{sec:lexcuts}

Another option, still based on the availability of the lexicographic simplex method, is a pure cutting plane method. Being a pure integer model, it is well-known that Gomory  cuts, together with lexicographic simplex, yield a cutting plane method converging in a finite number of iterations~\cite{Gomory58,Gomory60}. Recent computational studies show that the method can indeed converge in practice on some nontrivial models~\cite{BFZ10,ZFB11,BFZ12}. Unfortunately, a preliminary implementation of the method proved to be numerically unstable on our $n$-queens models.

However, it turns out that we can obtain a convergent method by using a different family of cutting planes, which we call \emph{lexicographic nogoods}, and that we now describe. Let $x^*$ be the optimal solution obtained by the lexicographic simplex method at the current iteration. If $x^*$ is integer, then we are done, otherwise let $x_{i^*j^*}$ be the first variable with a fractional value, i.e., $0 < x_{i^*j^*} < 1$. Finally, let $F$ be the (possibly empty) set of variables that precede $x_{i^*j^*}$ and that are assigned a value of $1$ in $x^*$. Then we can add the following cutting plane to the model:

\begin{align}
\sum_{(i,j) \in F} x_{ij} + x_{i^*j^*} \le |F|. \label{eq:lexnogood}
\end{align}

Note that, by definition, $F$ contains exactly one variable for each row $i < i^*$. The rationale behind the cut is as follows: $x^*$ being a lexicographically optimal solution, if we leave all variables in $F$ set to one, then we must set $x_{i^*j^*} = 0$. Otherwise we must flip at least one of the variables in $F$ to zero. In both cases \matt{inequality} (\ref{eq:lexnogood}) is valid and \matt{cuts} the fractional solution \matt{$x^*$}.

It is easy to show that the family of cuts (\ref{eq:lexnogood}), together with the lexicographic simplex, yields a convergent method: each cut forces \matt{a strict worsening of} to the lexicographic objective, thus the lexicographic simplex cannot cycle. \matt{As there is only a finite (albeit exponential) number of cuts, the} process must terminate in a finite number of iterations.

The pure cutting plane method based on cuts (\ref{eq:lexnogood}) did not eventually yield a faster algorithm than \texttt{LEX-DFS} in preliminary experiments, so we will not present it in the computational section. Still, it was able to solve almost as many chessboards as \texttt{LEX-DFS}, which is still remarkable for a pure cutting plane method.
}

\section{Computational comparisons}
\label{sec:comp}

We implemented our ILP models with the MIP solver \cplex\ \cite{cplex}, while we used Gecode~5.1.0~\cite{Gecode} as the CP solver for model (\ref{eq:alldiff1})-(\ref{eq:alldiff3}). All experiments were done on a cluster of 24 identical machines, each equipped with an Intel Xeon E3-1220 V2 quad-core PC and 16GB of RAM.

The testbed is made of all instances with $n$ ranging from $21$ to $60$. A time limit of 2 days was given for each instance to each method. Detailed results are given in Table~\ref{tab:rescomp}, where we report the running time, in seconds, for all of our methods.  The last two rows of the table report the shifted geometric mean~\cite{Achterberg07} of the computing time (with a shift of 10 sec.s) and the number of solved instances. According to the table, the CP model is able to solve models up to size 40 in a reasonable amount of time, after which it can no longer solve any model. Comparing with the numbers reported in~\cite{web}, this can be already considered a good achievement, and a testament to how efficient Gecode's implementation is. On the other hand, all methods based on ILP, while initially slower, turn out to be able to solve almost all models in the testbed. Among the ILP methods, \texttt{ILP-ITER}, while being the easiest to implement, is also the slowest method, while \texttt{ILP-TRUNC} and \texttt{LEX-DFS} are the fastest methods, with very similar average running times.

As already noted in~\cite{web}, the size of the chessboard is not a direct indicator of instance difficulty, as some bigger chessboards can be solved significantly faster than smaller ones. This is true in particular for ILP-based methods, where for example $n=48$ is unsolved while $n=49$ can be cracked in a few seconds. Interestingly, chessboards with even $n$ seem to be consistently harder than the ones with odd $n$.

As for the advanced techniques implemented in \texttt{LEX-DFS}, we have to admit that for some of them the overall effect was rather disappointing. In particular, the separation of clique and odd-cycle inequalities, while able to reduce the number of enumerated nodes by more than a factor of 2, does not lead to a faster algorithm overall. To the contrary, disabling cut separation leads to a slightly faster method with an average runtime of 246 sec.s. Note that this is not due to the complexity of separating cuts, separation being extremely fast for both classes of inequalities, but rather for the reduced node throughput.

\def\tl{\emph{t.l.}}

\afterpage{
\begin{table}[!p]
\begin{center}
\begin{tabular}{rR{2cm}R{2cm}R{2cm}R{2cm}}
\toprule
& \multicolumn{4}{c}{methods} \\
\cmidrule{2-5}
$n$ & \texttt{CP} & \texttt{ILP-ITER} & \texttt{ILP-TRUNC} & \texttt{LEX-DFS} \\
\midrule
      21 &             0.01 &             0.30 &             0.45 &             0.08 \\
      22 &             0.95 &             1.63 &            16.67 &             9.20 \\
      23 &             0.02 &             0.40 &             0.60 &             0.11 \\
      24 &             0.20 &             0.60 &             2.95 &             0.82 \\
      25 &             0.03 &             0.49 &             0.79 &             0.12 \\
      26 &             0.16 &             0.84 &             1.59 &             0.42 \\
      27 &             0.17 &             0.59 &             0.90 &             0.09 \\
      28 &             0.84 &             1.13 &             2.08 &             1.06 \\
      29 &             0.39 &             1.05 &             1.36 &             0.32 \\
      30 &            15.80 &            13.16 &            77.05 &            16.35 \\
      31 &             2.86 &             1.50 &             3.31 &             0.87 \\
      32 &            19.45 &             4.97 &            42.73 &             5.23 \\
      33 &            29.82 &            28.47 &            56.76 &            13.83 \\
      34 &           593.60 &           342.32 &          4558.02 &           228.07 \\
      35 &            33.70 &            11.21 &            30.46 &             4.67 \\
      36 &          5199.27 &          1882.10 &         20901.43 &          1196.59 \\
      37 &           185.37 &             2.06 &             7.49 &             0.54 \\
      38 &          2485.20 &           101.30 &           151.86 &           130.16 \\
      39 &          1642.30 &           143.02 &           184.50 &            44.79 \\
      40 & \tl              &          9604.18 &        117591.20 &          7068.84 \\
      41 &          1543.84 &            20.91 &           105.47 &             5.69 \\
      42 &  \tl             &  \tl             &  \tl             &  \tl             \\
      43 &         23528.50 &            21.65 &           162.13 &             6.08 \\
      44 & \tl              &          1013.43 &         14838.95 &          2220.52 \\
      45 & \tl              &          3604.37 &          4560.69 &          1388.93 \\
      46 & \tl              &  \tl             &  \tl             &  \tl             \\
      47 & \tl              &          1602.10 &          5057.63 &           601.54 \\
      48 & \tl              &  \tl             &  \tl             &  \tl             \\
      49 & \tl              &            23.26 &           460.07 &            10.35 \\
      50 & \tl              &         28011.44 &  \tl             &         61679.70 \\
      51 & \tl              &             4.63 &           874.16 &             0.88 \\
      52 & \tl              &         30306.67 &         27701.60 &         75659.40 \\
      53 & \tl              &             5.05 &           285.65 &             0.96 \\
      54 & \tl              &            64.09 &         19784.91 &         67031.40 \\
      55 & \tl              &            44.50 &           569.58 &            18.42 \\
      56 & \tl              &         28026.57 &         13386.85 &        101936.00 \\
      57 & \tl              &            10.03 &          3961.54 &             5.87 \\
      58 & \tl              &  \tl             &        129596.69 &  \tl             \\
      59 & \tl              &         49647.30 &  \tl             &         18795.70 \\
      60 & \tl              &  \tl             &         39143.49 &  \tl             \\
\midrule
shmean   &   2945.76 &    780.20 &    251.85 &    263.79 \\
\#solved &        21 &        35 &        35 &        35 \\
\bottomrule
~ 
\end{tabular}

\caption{Comparison of different methods for $n=21,\ldots,60$, with a time limit of 172800 sec.s (2 days).}
\label{tab:rescomp}
\end{center}
\end{table}
\clearpage
} 

\matteo{

\section{Most-beautiful queens}
\label{sec:most-beautiful}      

In the most-beautiful version of the $n$-queens problem, each blackboard cell $(i,j)$ has a {\em cost} defined as $$d_{ij} = (2i-n-1)^2+(2j-n-1)^2$$ that gives (4 times the squared) distance of cell $(i,j)$ from the center of the blackboard, for $i,j=1,\dots,n$.

Let us define the {\em fingerprint}  of a feasible solution $x$ as the list $$\phi(x) = (d_{ij}: x_{ij}=1)$$ sorted in non-increasing order w.r.t. the given input costs $d_{ij}$. Note that the list can contain repeated (consecutive) entries if the cost coefficients are not unique. 
E.g., for $n$=6 the feasible solution $x$ with $x_{14}=x_{21}=x_{35}=x_{42}=x_{56}=x_{63}=1$ has $d_{14}=26$, $d_{21}=34$, $d_{35}=10$, $d_{42}=10$, $d_{56}=34$ and $d_{63}=26$, hence its fingerprint is $\phi(x)=(34,34,26,26,10,10)$; see Figure~\ref{fig:mostbeautiful} for an illustration.

The most-beautiful $n$-queens problem then  calls for a (not necessarily unique) solution $x$ whose fingerprint $\phi(x)$ is lexicographically minimal. 

This is a rather general setting, asking for a ``lexicographically bottleneck'' (or lexicographically min-max) optimal solution of a given combinatorial problem, hence the solution approach we propose in what follows extends to more general contexts.

\begin{figure}
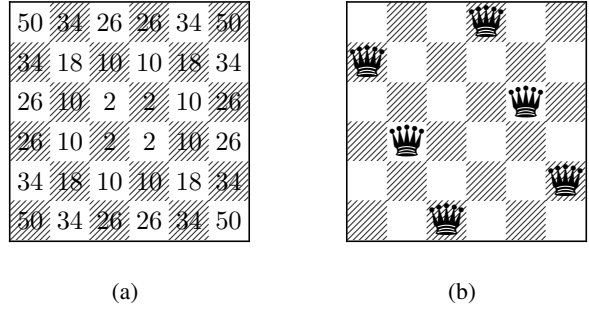

\begin{center}
\subfigure[]{
\chessboard[style=6x6,
	smallboard,
	color=black,
	pgfstyle=text,text=\fontsize{1.5ex}{1.5ex}\bfseries $50$,markfields={a6},
	pgfstyle=text,text=\fontsize{1.5ex}{1.5ex}\bfseries $34$,markfields={b6},
	pgfstyle=text,text=\fontsize{1.5ex}{1.5ex}\bfseries $26$, markfields={c6},
	pgfstyle=text,text=\fontsize{1.5ex}{1.5ex}\bfseries $26$, markfields={d6},
	pgfstyle=text,text=\fontsize{1.5ex}{1.5ex}\bfseries $34$, markfields={e6},
	pgfstyle=text,text=\fontsize{1.5ex}{1.5ex}\bfseries $50$, markfields={f6},
	pgfstyle=text,text=\fontsize{1.5ex}{1.5ex}\bfseries $34$, markfields={a5},
	pgfstyle=text,text=\fontsize{1.5ex}{1.5ex}\bfseries $18$, markfields={b5},
	pgfstyle=text,text=\fontsize{1.5ex}{1.5ex}\bfseries $10$, markfields={c5},
	pgfstyle=text,text=\fontsize{1.5ex}{1.5ex}\bfseries $10$, markfields={d5},
	pgfstyle=text,text=\fontsize{1.5ex}{1.5ex}\bfseries $18$, markfields={e5},
	pgfstyle=text,text=\fontsize{1.5ex}{1.5ex}\bfseries $34$, markfields={f5},
	pgfstyle=text,text=\fontsize{1.5ex}{1.5ex}\bfseries $26$, markfields={a4},
	pgfstyle=text,text=\fontsize{1.5ex}{1.5ex}\bfseries $10$, markfields={b4},
	pgfstyle=text,text=\fontsize{1.5ex}{1.5ex}\bfseries $ 2$, markfields={c4},
	pgfstyle=text,text=\fontsize{1.5ex}{1.5ex}\bfseries $ 2$, markfields={d4},
	pgfstyle=text,text=\fontsize{1.5ex}{1.5ex}\bfseries $10$, markfields={e4},
	pgfstyle=text,text=\fontsize{1.5ex}{1.5ex}\bfseries $26$, markfields={f4},
	pgfstyle=text,text=\fontsize{1.5ex}{1.5ex}\bfseries $26$, markfields={a3},
	pgfstyle=text,text=\fontsize{1.5ex}{1.5ex}\bfseries $10$, markfields={b3},
	pgfstyle=text,text=\fontsize{1.5ex}{1.5ex}\bfseries $ 2$, markfields={c3},
	pgfstyle=text,text=\fontsize{1.5ex}{1.5ex}\bfseries $ 2$, markfields={d3},
	pgfstyle=text,text=\fontsize{1.5ex}{1.5ex}\bfseries $10$, markfields={e3},
	pgfstyle=text,text=\fontsize{1.5ex}{1.5ex}\bfseries $26$, markfields={f3},
	pgfstyle=text,text=\fontsize{1.5ex}{1.5ex}\bfseries $34$, markfields={a2},
	pgfstyle=text,text=\fontsize{1.5ex}{1.5ex}\bfseries $18$, markfields={b2},
	pgfstyle=text,text=\fontsize{1.5ex}{1.5ex}\bfseries $10$, markfields={c2},
	pgfstyle=text,text=\fontsize{1.5ex}{1.5ex}\bfseries $10$, markfields={d2},
	pgfstyle=text,text=\fontsize{1.5ex}{1.5ex}\bfseries $18$, markfields={e2},
	pgfstyle=text,text=\fontsize{1.5ex}{1.5ex}\bfseries $34$, markfields={f2},
	pgfstyle=text,text=\fontsize{1.5ex}{1.5ex}\bfseries $50$, markfields={a1},
	pgfstyle=text,text=\fontsize{1.5ex}{1.5ex}\bfseries $34$, markfields={b1},
	pgfstyle=text,text=\fontsize{1.5ex}{1.5ex}\bfseries $26$, markfields={c1},
	pgfstyle=text,text=\fontsize{1.5ex}{1.5ex}\bfseries $26$, markfields={d1},
	pgfstyle=text,text=\fontsize{1.5ex}{1.5ex}\bfseries $34$, markfields={e1},
	pgfstyle=text,text=\fontsize{1.5ex}{1.5ex}\bfseries $50$, markfields={f1},
	addpgf]
\label{fig:beautyobj}}
\subfigure[]{
\chessboard[style=6x6,
	smallboard,
	setblack={Qa5,Qb3,Qc1,Qd6,Qe4,Qf2}]
\label{fig:beautysol}}
\caption{Most beautiful $n$-queens costs and a feasible (actually, optimal) arrangement for $n=6$,
with fingerprint $(34,34,26,26,10,10)$.}
\label{fig:mostbeautiful}
\end{center}
\end{figure}

When stated in the above way, the most-beautiful $n$-queens problem can be solved as in Algorithm~\ref{alg:beauty}, where the different cost values $D_k$ (say) are scanned in decreasing order, and an ILP model is solved to find (and then fix) the minimum number of selected cells $(i,j)$ sharing the same cost $d_{ij}=D_k$.

\begin{algorithm}
\LinesNumbered
\BlankLine
{
	build an ILP model for the standard $n$-queens problem, with no objective function\;
	sort the costs $d_{ij}$'s in decreasing order (removing duplicates) and obtain the list of $m$ (say) distinct costs $D_1 > D_2 > \dots > D_m$\;
	\For{$k=1,\dots,m$} 	
	{
			solve the current ILP model with objective function $\sum_{i,j: d_{ij}=D_k} x_{ij}$ (to be minimized), and let $x^*$ be the optimal solution found and $z^*$ its value\; \label{Step:For1}
			add the constraint $\sum_{i,j: d_{ij}=D_k} x_{ij} = z^*$ to the current ILP model  \label{Step:For2}
	}
	\Return{$x^*$}
}
\caption{ILP-based solver for the most-beautiful $n$-queens problem.}\label{alg:beauty}
\end{algorithm}

At Step~\ref{Step:For2}, in case $z^*=0$ one can more conveniently set $x_{ij}=0$ for all $(i,j)$ such that $d_{ij}=D_k$. Note however that, when $z^*>0$, one {\em cannot} fix $x_{ij}=1$ for $d_{ij}=D_k$ and $x^*_{ij}=1$, as solution $x^*$ is not necessarily unique---hence this fixing would affect the correctness of the algorithm.  In addition, one can exit the for-loop as soon as the sum of the right-hand-side values $z^*$ of the cardinality constraints added at Step \ref{Step:For2} reaches $n$.

We observed that most iterations (in particular, the first ones) produce $z^*=0$; e.g., for $n=128$ this occurs for the first 397 (out of 1464) iterations. In this situation, the computing time of the overall algorithm is highly affected by the availability of heuristics that are able to find very quickly a solution not using any cell $(i,j)$ with $d_{ij}=D_k$---possibly starting from the optimal solution found at the previous iteration. Modern ILP solvers do have such parametric heuristics in their arsenal, which is highly beneficial for the overall computing time.

To gain an additional speedup, we implemented the following simple preprocessing mechanism: We start with $k=1$ and try to find a feasible ILP solution $x^*$ with $x^*_{ij}=0$ for all $(i,j)$ such that $d_{ij}\ge D_{k+100}$. (To abort the ILP solver as soon as possible, we provide a very small upper cutoff on input to the ILP solver, namely 0.01 in our implementation). If we are successful, we fix to zero all those $x_{ij}$ variables, increase $k$ by 100, and repeat. Otherwise, we just enter the for-loop at Step~\ref{Step:For1} with the current value of $k$.

} 

\section{Conclusions and future directions of work}
\label{sec:conclusions}

Finding a lexicographically minimal (also called ``first'') solution of the $n$-queens puzzle is a very difficult problem that attracted some research interest in recent years.  Following a suggestion by Donald E. Knuth, we have developed new solution methods based on Integer Linear Programming, and have been able to provide the optimal solution for several open problems.

The two main outcomes of our research are as follows: (1) ILP has been able to solve many previously unsolved models for this problem, sometimes in unexpectedly-short computing times; (2) the yet-unsolved cases provide excellent benchmark examples on which to base the next advances in ILP technology. In addition, we think that improving our understanding on how to solve lexicographic variants of combinatorial problems is an interesting topic on its own. \domenico{Finally, we developed a convergent pure cutting plane method \matt{based on a combinatorial variant of Gomory cuts that we called \emph{lexicographic nogood} cuts. Though not competitive in our case, this method is} theoretically interesting and can be possibly extended to more general binary integer programs.}

\matteo{We also addressed the ``most beautiful'' version of the problem, that calls for a lexicographically bottleneck (or min-max) solution, and proposed a new ILP-based solution scheme capable of discovering those solutions for some open cases.}

Future research should address the unsolved cases, and in particular should try to better understand the reason why, \matteo{for the lexicographically-first version, the ILP instances} with even $n$ seem to be much more difficult to solve than those with $n$ odd.

\section*{Acknowledgements}                           
This research was partially supported by MiUR, Italy, through project PRIN2015 ``Nonlinear and Combinatorial Aspects of Complex Networks''. 
\matte{The research of the first author was also supported by the Vienna Science and Technology Fund (WWTF) through project ICT15-014.}. 
We thank Donald E. Knuth for having pointed out the problem to us, and for inspiring discussions on the role of Integer Linear Programming in solving combinatorial problems arising in digital tomography.
\label{sec:ack}


\section*{Appendix: Solutions}
\label{sec:record}

Here are the lexicographically-first solutions we found for some open problems from the literature:

\begin{center}
\begin{longtable}{L{1cm}p{9cm}}
\toprule
$n$ & \matteo{Lexicographically-first solution} \\
\midrule
\endfirsthead
\toprule
$n$ & \matteo{Lexicographically-first solution} \\
\midrule
\endhead

\midrule
\multicolumn{2}{r}{{(continued on next page)}} \\
\endfoot

\bottomrule
\endlastfoot

56 & 1 3 5 2 4 9 11 13 15 6 8 19 7 22 10 25 27 29 31 33 42 44 46 43 51 53 55 45 54 50 47 56 48 52 49 12 14 23 21 32 34 26 16 30 17 24 18 37 28 40 20 39 41 35 38 36 \\
57 & 1 3 5 2 4 9 11 13 15 6 8 19 7 22 10 25 27 29 31 12 34 43 45 47 50 52 54 44 57 49 46 56 51 48 55 53 14 28 17 33 23 16 18 30 24 37 20 32 21 26 40 35 41 39 42 36 38 \\
58 & 1 3 5 2 4 9 11 13 15 6 8 19 7 22 10 25 27 29 31 12 42 45 48 52 54 43 53 55 49 44 46 50 57 47 51 58 56 28 26 20 34 30 18 14 17 24 21 16 35 23 40 33 36 38 32 41 39 37 \\
59 & 1 3 5 2 4 9 11 13 15 6 8 19 7 22 10 25 27 29 31 12 34 36 45 47 49 52 56 53 46 57 59 48 51 54 50 55 58 16 14 17 32 23 26 20 18 33 35 28 21 43 41 37 24 40 44 30 39 42 38 \\
60 & 1 3 5 2 4 9 11 13 15 6 8 19 7 22 10 25 27 29 31 12 34 44 46 48 45 51 54 58 50 59 57 60 47 49 52 55 53 56 18 33 23 32 28 16 20 17 21 37 35 26 24 30 14 42 38 43 41 39 36 40 \\
61 & 1 3 5 2 4 9 11 13 15 6 8 19 7 22 10 25 27 29 31 12 14 35 45 47 49 52 54 56 50 60 46 61 58 48 51 53 55 57 59 23 32 16 33 21 17 26 36 18 20 38 24 28 34 40 30 41 44 42 37 39 43 \\
63 & 1 3 5 2 4 9 11 13 15 6 8 19 7 22 10 25 27 29 31 12 14 35 37 47 49 51 53 59 57 52 60 62 48 50 54 63 55 58 56 61 32 16 33 17 21 26 36 20 18 38 28 23 40 24 30 34 41 39 44 46 43 45 42 \\
65 & 1 3 5 2 4 9 11 13 15 6 8 19 7 22 10 25 27 29 31 12 14 35 37 39 49 51 53 50 56 59 63 55 64 62 65 52 54 57 60 58 61 16 30 17 21 26 36 33 20 18 41 38 23 32 24 28 48 46 34 43 40 44 47 45 42 \\
67 & 1 3 5 2 4 9 11 13 15 6 8 19 7 22 10 25 27 29 31 12 14 35 37 39 41 51 53 55 52 58 61 65 57 66 64 67 54 56 59 62 60 63 16 18 34 30 38 20 24 17 21 23 43 32 40 33 36 26 28 46 48 50 44 47 45 42 49 \\
69 & 1 3 5 2 4 9 11 13 15 6 8 19 7 22 10 25 27 29 31 12 14 35 37 39 41 43 53 55 57 54 60 63 67 59 68 66 69 56 58 61 64 62 65 17 20 16 30 24 33 40 38 18 21 34 26 23 42 49 28 32 50 36 51 46 44 52 48 45 47 \\
71 & 1 3 5 2 4 9 11 13 15 6 8 19 7 22 10 25 27 29 31 12 14 35 37 39 41 16 53 55 57 54 56 62 68 66 69 59 70 67 58 71 61 64 60 65 63 21 30 17 40 18 24 36 20 42 44 26 34 23 33 38 32 28 49 51 45 47 52 50 48 46 43 \\
73 & 1 3 5 2 4 9 11 13 15 6 8 19 7 22 10 25 27 29 31 12 14 35 37 39 41 16 44 55 57 59 56 58 63 67 69 71 73 61 70 72 65 60 62 64 66 68 20 34 21 18 42 17 38 24 43 23 28 45 33 40 36 26 32 30 54 47 50 52 46 48 53 51 49 \\
77 & 1 3 5 2 4 9 11 13 15 6 8 19 7 22 10 25 27 29 31 12 14 35 37 39 41 16 18 45 57 59 61 58 60 65 68 72 74 76 73 75 63 67 64 62 77 70 66 71 69 38 40 28 17 21 24 26 20 43 46 42 23 36 34 32 30 44 33 52 55 47 50 53 56 54 48 51 49 \\
79 & 1 3 5 2 4 9 11 13 15 6 8 19 7 22 10 25 27 29 31 12 14 35 37 39 41 16 18 45 47 59 61 63 60 62 67 70 74 71 77 79 76 78 64 68 65 69 66 73 75 72 20 38 17 21 44 24 30 23 46 48 36 42 40 34 26 28 33 50 32 53 43 57 52 58 56 54 51 49 55 \\
85 & 1 3 5 2 4 9 11 13 15 6 8 19 7 22 10 25 27 29 31 12 14 35 37 39 41 16 18 45 17 48 50 63 65 67 64 66 71 73 75 80 82 84 81 83 72 70 68 85 69 78 74 77 79 76 20 23 43 24 21 49 44 42 34 46 28 30 52 26 38 51 32 40 33 61 47 60 36 53 58 54 57 59 56 62 55 \\
91 & 1 3 5 2 4 9 11 13 15 6 8 19 7 22 10 25 27 29 31 12 14 35 37 39 41 16 18 45 17 48 20 51 53 67 69 71 68 70 75 77 79 81 85 87 90 86 91 89 72 74 76 73 80 82 84 78 83 88 21 34 26 49 46 24 47 52 43 23 30 33 55 28 42 32 54 40 36 44 64 50 38 59 61 65 57 66 60 63 56 58 62 \\
93 & 1 3 5 2 4 9 11 13 15 6 8 19 7 22 10 25 27 29 31 12 14 35 37 39 41 16 18 45 17 48 20 51 53 55 69 71 73 70 72 77 79 81 83 87 89 92 88 93 91 74 76 78 75 82 84 86 80 85 90 24 21 23 46 49 47 52 38 30 56 33 26 28 43 32 54 57 42 44 36 34 40 50 61 68 65 62 59 63 58 67 64 66 60 \\
97 & 1 3 5 2 4 9 11 13 15 6 8 19 7 22 10 25 27 29 31 12 14 35 37 39 41 16 18 45 17 48 20 51 53 21 56 71 73 75 72 74 79 81 83 85 87 89 93 95 97 94 96 76 80 77 86 78 82 84 91 88 90 92 46 24 28 52 23 49 47 34 30 26 57 50 33 61 42 44 36 32 55 43 38 54 60 66 40 70 68 63 58 69 62 65 67 64 59 \\
101 & 1 3 5 2 4 9 11 13 15 6 8 19 7 22 10 25 27 29 31 12 14 35 37 39 41 16 18 45 17 48 20 51 53 21 56 58 60 75 77 79 76 78 83 85 87 89 91 93 97 99 101 98 100 80 84 81 90 82 86 88 95 92 94 96 23 26 28 40 43 54 57 24 32 47 50 42 59 33 30 34 52 62 68 46 38 36 44 55 66 71 74 70 49 73 63 72 67 61 64 69 65 \\
103 & 1 3 5 2 4 9 11 13 15 6 8 19 7 22 10 25 27 29 31 12 14 35 37 39 41 16 18 45 17 48 20 51 53 21 56 58 60 62 77 79 81 78 80 85 87 89 91 93 95 99 101 103 100 102 82 86 83 92 84 88 90 97 94 96 98 23 26 24 30 28 36 46 55 59 52 54 44 61 34 66 33 42 32 47 49 40 38 57 73 71 63 72 43 64 70 75 50 69 67 76 74 68 65 \\
109 & 1 3 5 2 4 9 11 13 15 6 8 19 7 22 10 25 27 29 31 12 14 35 37 39 41 16 18 45 17 48 20 51 53 21 56 58 60 23 63 65 81 83 85 82 84 89 91 93 95 86 100 104 106 101 109 107 105 108 88 92 87 96 90 97 102 94 98 103 99 26 24 32 28 36 55 57 40 64 61 54 50 30 66 34 42 38 33 49 43 67 59 62 77 52 44 47 75 71 46 76 80 73 70 79 69 78 72 74 68 \\
115 & 1 3 5 2 4 9 11 13 15 6 8 19 7 22 10 25 27 29 31 12 14 35 37 39 41 16 18 45 17 48 20 51 53 21 56 58 60 23 63 24 66 68 85 87 89 86 88 93 95 97 99 90 102 108 111 113 107 109 112 115 91 114 98 101 92 94 96 100 105 103 110 106 104 26 28 30 32 36 50 59 62 64 55 43 34 72 67 52 33 40 65 57 44 42 38 74 54 61 46 83 47 77 69 49 82 79 75 84 71 80 78 81 73 70 76 \\
\end{longtable}
\end{center}

\matteo{

And these are some most-beautiful solutions found by our ILP-based approach (those with $n > 48$ are new); the reported computing times are wall-clock seconds on a notebook (Intel Core i7 \@2.3GHz with 16GB RAM).

\begin{center}
\begin{longtable}{R{1cm}R{1.2cm}p{9cm}}
\toprule
$n$ & time (sec.s) & \matteo{Most-beautiful solution} \\
\midrule
\endfirsthead
\toprule
$n$ & time (sec.s) & \matteo{Most-beautiful solution} \\
\midrule
\endhead

\midrule
\multicolumn{3}{r}{{(continued on next page)}} \\
\endfoot

\bottomrule
\endlastfoot
16  &       0.2  & 9 11 4 14 10 2 5 1 16 12 15 7 3 13 6 8 \\
32  &       2.2  & 17 14 12 23 18 26 6 11 13 4 25 30 24 31 5 1 32 28 2 9 3 8 29 20 22 27 7 15 10 21 19 16 \\  
48  &      21.7  & 25 27 29 26 17 34 12 38 9 18 30 21 7 43 36 5 14 45 37 3 6 2 39 1 48 41 47 42 46 8 4 11 44 15 19 32 28 33 40 10 23 13 35 16 31 20 22 24  \\
64  &      43.0  & 33 35 37 34 40 42 36 20 17 48 13 52 12 27 38 55 43 39 8 24 58 44 6 15 5 21 51 3 7 2 4 1 64 61 63 50 62 56 16 60 47 59 22 18 46 57 9 41 10 26 54 11 53 14 49 19 45 29 23 25 31 28 30 32  \\
80  &     137.8  & 41 43 36 42 48 31 44 28 55 57 59 35 18 65 15 66 49 47 68 12 52 11 29 10 51 9 23 73 56 64 75 17 5 21 19 78 7 2 4 1 80 77 79 74 3 62 60 76 61 6 20 25 8 58 72 30 71 26 70 53 69 13 34 14 67 32 16 63 46 22 24 54 27 37 50 33 39 45 38 40 \\
96  &     264.6  & 49 51 44 50 56 59 36 34 32 62 68 57 26 45 74 22 20 18 80 37 81 55 39 58 67 84 28 66 11 70 27 88 30 89 10 90 72 6 25 12 5 79 82 94 15 2 4 1 96 93 95 14 3 21 77 92 85 73 91 23 7 87 8 75 9 31 69 86 24 64 13 33 83 40 16 54 60 17 19 78 76 42 52 71 43 29 35 65 63 61 38 41 47 53 46 48  \\
112  &     830.5  & 57 59 61 58 49 47 68 70 72 74 37 65 34 81 48 84 62 27 89 22 92 21 53 46 19 95 44 71 36 38 42 80 13 100 39 31 102 77 10 82 9 88 8 16 7 86 107 98 108 97 15 3 90 2 4 1 112 109 111 23 110 96 17 5 101 6 25 106 29 105 26 104 30 103 11 28 12 78 40 83 99 14 33 75 67 69 18 73 94 60 20 93 91 24 87 85 63 51 32 79 35 76 50 41 43 45 66 64 55 52 54 56  \\
128  &    3181.7  & 64 67 69 63 57 74 68 77 50 81 83 70 86 41 39 37 54 34 59 98 101 26 105 23 25 107 78 21 108 49 53 19 76 111 88 51 113 93 114 43 115 32 13 94 87 11 29 10 97 120 35 8 99 27 6 112 124 15 104 3 7 2 125 128 1 4 127 122 126 117 18 5 17 123 100 96 121 91 9 30 119 89 118 42 12 38 116 14 95 45 36 16 84 47 44 110 82 109 20 80 75 56 22 106 103 24 102 28 31 33 71 73 92 90 40 58 85 46 48 79 52 61 55 72 66 60 62 65  \\
144  &    2933.0  & 73 75 77 74 80 82 76 60 87 89 91 78 50 97 47 100 102 104 106 81 79 86 34 114 29 27 119 117 120 84 24 62 23 90 123 59 52 101 19 42 18 39 17 53 16 40 95 103 131 48 13 133 112 11 37 10 107 9 32 137 108 115 6 20 5 21 12 3 7 2 4 1 144 141 143 138 142 28 130 140 125 139 128 124 8 113 136 35 135 111 134 44 99 132 109 14 46 15 36 129 92 38 127 57 93 126 88 49 94 22 55 122 83 121 61 25 30 26 118 116 31 33 110 66 64 105 41 43 45 98 96 51 67 54 56 58 85 69 63 65 71 68 70 72 \\
160  &   38757.7  & 81 83 85 82 88 90 84 93 95 97 99 101 74 57 106 53 110 112 47 91 69 67 72 121 38 36 127 31 29 132 33 133 27 75 26 135 100 136 60 52 137 23 54 102 107 105 142 115 18 103 17 118 145 46 15 111 147 45 13 124 44 11 119 10 41 9 35 8 141 139 6 22 5 126 149 3 7 2 4 1 160 157 159 154 158 12 21 156 128 155 24 20 153 42 152 123 151 122 150 116 39 148 120 14 50 146 43 16 48 144 58 143 117 19 113 109 140 96 61 138 65 63 25 98 56 87 134 32 130 28 30 131 129 34 125 37 40 89 94 92 70 114 49 51 108 55 104 59 86 62 64 66 68 77 71 73 79 76 78 80 \\
176  &   10758.2  & 89 91 93 90 81 98 85 101 103 105 107 109 66 113 115 60 119 57 123 100 126 49 102 73 97 44 42 95 38 36 142 32 145 31 141 99 147 108 148 28 112 150 63 67 152 71 122 106 59 48 21 129 61 19 158 18 46 132 120 161 41 15 47 14 40 164 136 12 133 11 137 10 155 9 20 8 23 35 6 39 172 25 143 3 170 2 4 1 176 173 175 7 174 149 37 5 151 171 17 157 169 154 168 24 167 45 166 139 165 134 13 128 163 121 162 52 16 43 160 131 159 116 118 50 127 156 55 22 125 53 153 114 26 65 27 69 77 29 110 30 78 146 94 33 144 34 140 138 83 135 80 82 130 75 104 51 124 54 56 58 117 62 64 111 68 70 72 74 76 92 79 96 87 84 86 88 
\end{longtable}
\end{center}

} 

\bibliography{queens}
\bibliographystyle{plain}

\end{document}